  \documentclass{gretsi}
\usepackage[latin1]{inputenc}   
  \usepackage{times}			
 
\usepackage[T1]{fontenc}

\usepackage{amsmath,graphicx}
\usepackage{amsfonts,amssymb,amsthm}
\usepackage{mathrsfs}
\usepackage[Algorithm,ruled]{algorithm}
\usepackage{algorithmic}
\usepackage{xcolor}
\usepackage{multirow}
\usepackage{hhline}
 \usepackage[english,french]{babel}
\def\bz{{\mathbf z}}
\def\bA{{\mathbf A}}
\def\bB{{\mathbf B}}

\def\bM{{\mathbf M}}
\def\by{{\mathbf y}}
\def\bs{{\mathbf s}}
\def\bw{{\mathbf w}}

\def\ccS{\mathscr{S}}
\def\cD{\mathcal{D}}
\def\cW{\mathcal{W}}
\def\RR{\mathbb{R}}
\def\CC{\mathbb{C}}
\def\EE{\mathbb{E}}
\def\btheta{\boldsymbol{\theta}}
\def\bSigma{\boldsymbol{\Sigma}}

\def\bepsilon{\boldsymbol{\epsilon}}
\def\bgamma{\boldsymbol{\gamma}}
\def\defeq{\stackrel{\Delta}{=}}

\newtheorem{theorem}{ThÈorËme}

\titre{SÈparation de sources doublement non stationnaire}

   \auteur{\coord{Adrien}{Meynard}{}}                                       %
   \adresse{\affil{}{Aix Marseille Univ, CNRS, Centrale Marseille, I2M, Marseille, France}}   %
\email{adrien.meynard@univ-amu.fr}

\resumefrancais{Les techniques de sÈparation aveugle de sources (\textit{blind source separation}, BSS) visent ‡ estimer conjointement des signaux sources et une matrice de mÈlange ‡ partir des observations de mÈlanges. Cet article aborde un problËme de BSS doublement non stationnaire, o˘ la matrice de mÈlange dÈpend du temps et les sources sont non stationnaires. Plus prÈcisÈment, ce sont des signaux stationnaires dÈformÈs, suivant le modËle de~\cite{Meynard18spectral}. Nous prÈsentons un algorithme joint pour la BSS et l'estimation des dÈformations brisant la stationnaritÈ et des spectres. Il s'appuie sur des approximations appropriÈes du comportement de la transformÈe en ondelettes de ces signaux non stationnaires. Les performances cet algorithme sont ÈvaluÈes sur des simulations numÈriques et comparÈes ‡ d'autres algorithmes de BSS non stationnaires.}

\resumeanglais{Blind source separation (BSS) techniques aims at joint estimation of source signals and a mixing matrix from observations of mixtures. This paper addresses a doubly nonstationary BSS problem, where the mixing matrix is time dependent and sources are nonstationary, more precisely deformed stationary signals, following the model of~\cite{Meynard18spectral}. An algorithm for joint BSS and estimation of stationarity-breaking deformations and spectra is introduced, that exploits suitable approximations for the behavior of the wavelet transform of such nonstationary signals. The performance of the approach is evaluated on numerical simulations, and compared with other nonstationary BSS algorithms.}

\begin{document}
\maketitle

\section{Introduction}
\vspace{-3mm}
La plupart des algorithmes de BSS s'appuient sur une hypothËse de stationnaritÈ des sources ‡ estimer, et traitent souvent aussi des mÈlanges linÈaires instantanÈs o˘ la matrice de mÈlange est constante dans le temps (voir~\cite{Comon10handbook} pour une Ètude approfondie du sujet). Par exemple, SOBI~\cite{Belouchrani97blind} constitue une approche de rÈfÈrence qui exploite les statistiques de second ordre et dont l'estimation repose sur les diagonalisations approchÈes et jointes des matrices de covariance. Cependant, ces hypothËses sur le modËle de mÈlange limitent l'application de la BSS ‡ un petit nombre de signaux et de situations.

Plusieurs extensions ‡ la BSS de mÈlanges linÈaires instantanÈs de signaux non stationnaires ont ÈtÈ proposÈes. La plupart d'entre elles reposent sur une analyse temps-frÈquence quadratique (voir par ex.~\cite{Thirion10quadratic, Belouchrani13source}) et s'appuient sur une sÈlection de points du domaine temps-frÈquence pour lesquels une seule des sources est active (se rÈfÈrer ‡~\cite{Thirion10quadratic} pour plus de dÈtails sur la sÈlection des points dans le cas d'une distribution de Wigner modifiÈe). Nous mettrons en \oe{}uvre une de ces techniques pour comparaison, appelÈe QTF-BSS dans le document. Des algorithmes de BSS, fondÈs sur l'indÈpendance entre sources non stationnaires et l'indÈpendance temporelle de chaque source, ont Ègalement ÈtÈ proposÈs. Dans~\cite{Pham01blind}, l'idÈe est d'approcher les sources non stationnaires par des signaux stationnaires par morceaux. Ainsi, cette mÈthode s'approche d'une estimation SOBI par morceaux sur des sous-intervalles ne se chevauchant pas (elle est notÈe p-SOBI dans l'article).

En outre, afin de modÈliser des situations physiques (par ex.: rÈverbÈration de signaux audio), des modËles de mÈlange non instantanÈs ont ÈtÈ proposÈs. Dans~\cite{Kaftory13blind}, les auteurs considËrent le cas trËs gÈnÈral d'un mÈlange convolutif variant dans le temps (ce qui inclue le mÈlange instantanÈe non constant). NÈanmoins, la mÈthode de BSS associÈe repose sur l'hypothËse qu'il existe une reprÈsentation des sources dans laquelle elles sont parcimonieuses. Il s'agit alors, comme pour QTF-BSS, de sÈlectionner les points o˘ une seule source est active. Concernant la BSS d'un mÈlange linÈaire variant dans le temps, l'approche de diffÈrents algorithmes de BSS (voir par ex.~\cite{Prieto05blind}) consiste ‡ effectuer, ‡ l'instar de p-SOBI, une analyse en composantes indÈpendantes sur de courtes portions du mÈlange sur lesquelles la stationnaritÈ du mÈlange peut Ítre supposÈe.

Dans ce qui suit, nous nous intÈressons ‡ un problËme de BSS doublement non stationnaire et abordons le problËme des signaux non stationnaires, mÈlangÈs par une matrice de mÈlange non stationnaire. Sur la base de travaux antÈrieurs sur la non-stationnaritÈ et la dÈformation temporelle~\cite{Clerc03estimating, Omer17time}, nous nous concentrons sur une classe spÈcifique de signaux non stationnaires, ‡ savoir les signaux stationnaires composÈs avec une dÈformation temporelle. Dans~\cite{Meynard18spectral}, on analyse ce type de signaux individuellement. On Ètend ici cette analyse ‡ des mÈlanges de tels signaux, et nous montrons que si les matrices de mÈlange et les fonctions de dÈformation temporelle sont suffisamment lisses et lentement variables, la vraisemblance approchÈe sur les transformÈes en ondelettes des observations peut Ítre dÈduite sous une hypothËse de sources gaussiennes. Par consÈquent, l'estimation conjointe de la matrice de mÈlange et des fonctions de dÈformation temporelle constitue un problËme de maximum de vraisemblance. L'algorithme de BSS proposÈ consiste en une estimation alternÈe de la matrice de mÈlange et des fonctions de dÈformation. Cette derniËre estimation est effectuÈe par l'algorithme JEFAS (\textit{Joint Estimation of Frequency, Amplitude and Spectrum}) introduit dans nos travaux prÈcÈdents (voir~\cite{Meynard18spectral}, code MATLAB disponible en ligne \footnotemark[1]).

\footnotetext[1]{\texttt{https://github.com/AdMeynard/JEFAS}}

\section{ModËle}
\subsection{Une classe de signaux non stationnaires}
Dans ce qui suit, nous considÈrons les signaux non stationnaires comme des versions dÈformÈes de signaux stationnaires. Ces hypothËses sont appropriÈes pour dÈcrire de nombreux signaux de la vie rÈelle. En effet, diffÈrentes classes d'opÈrateurs qui brisent la stationnaritÈ sont pertinentes pour modÈliser des phÈnomËnes physiques (par ex. la modulation de frÈquence, ou la modulation d'amplitude~\cite{Meynard18spectral}). Nous nous concentrons ici sur l'opÈrateur de dÈformation temporelle. Un tel opÈrateur peut modÈliser des phÈnomËnes physiques non stationnaires aussi divers que l'effet Doppler, les variations de vitesse d'un moteur, la vocalisation d'animaux ou la parole~\cite{Stowell18computational, Meynard18spectral}.

Introduisons maintenant quelques notations. Soit $ x $ un signal stationnaire, modÈlisÈ comme la rÈalisation d'un processus alÈatoire stationnaire $ X $ dont le densitÈ spectrale de puissance est notÈe $ \ccS_X $. En agissant sur $ x $ avec l'opÈrateur de dÈformation temporelle notÈ $ \cD_\gamma $, on obtient un signal non stationnaire notÈ $ y $. La dÈformation temporelle est dÈfinie par:
\begin{equation}
\label{eq:nonstat.model}
y(t) = \cD_\gamma x (t)= \sqrt{\gamma'(t)}\,x(\gamma(t))\ ,
\end{equation}
o˘ $\gamma\in C^2$ est une fonction strictement croissante.

La transformÈe en ondelettes est un outil naturel pour analyser de tels signaux. \`A partir d'un ondelette complexe $\psi$, on dÈfinit la transformÈe en ondelettes $ \cW_x $ du signal $ x $ par:
\begin{equation}
\cW_x(s,\tau) = \int_\RR x(t)q^{-\frac{s}{2}}\overline{\psi}\left(\dfrac{t-\tau}{q^s}\right)dt\quad \text{avec}\quad q>1\ .
\end{equation}
Dans ce cadre, on peut montrer~\cite{Omer17time,Meynard18spectral} que les transformÈes en ondelettes respectives $\cW_y$ et $\cW_x$ de $y$ et  $x$ sont approximativement reliÈes par l'Èquation suivante:
\begin{equation}
\label{eq:approx.wavelet}
\cW_y(s,\tau) \approx \cW_x(s + \log_q(\gamma'(\tau)),\gamma(\tau))\ .
\end{equation}
Dans le cadre de processus alÈatoires, on peut montrer~\cite{Omer17time,Meynard18spectral} que l'erreur d'approximation est non biaisÈe et que sa variance peut Ítre contrÙlÈe gr‚ce aux propriÈtÈs de dÈcroissance de l'ondelette $ \psi $ et aux variations de $\gamma'$.

\subsection{MÈlange instantanÈ non stationnaire}
Le problËme que nous considÈrons est la BSS de signaux non stationnaires modÈlisÈs par l'Èquation~\eqref{eq:nonstat.model}. Les sources sont en outre supposÈes non corrÈlÈes. Nous supposons Ègalement que le nombre de sources $ M $ est Ègal au nombre de mÈlanges $ N $. Dans le cas surdÈterminÈ, on effectue classiquement une rÈduction de dimension pour revenir au cas $ M = N $. Le cas sous-dÈterminÈ dÈpasse le cadre du prÈsent travail.

Soit $ \by (t), \bz (t) \in \RR^N $ les vecteurs colonnes contenant respectivement toutes les sources et observations ‡ l'instant $ t $. Le mÈlange s'Ècrit alors
\begin{equation}
\label{eq:bss.model}
\bz(t) = \bA(t)\by(t)\ ,
\end{equation}
o˘ $\bA(t)\in\RR^{N \times N}$ dÈsigne la matrice de mÈlange variant dans le temps, supposÈe Ítre inversible. Ce modËle gÈnÈralise le modËle de modulation d'amplitude correspondant au cas o˘ $ \bA (t) $ est diagonale.

Notre objectif est de dÈterminer conjointement la matrice de mÈlange $ \bA (t) $, les fonctions de dÈformation temporelle $ \gamma_i (t) $ et le spectre des sources stationnaires $ \ccS_ {X_i} $ pour $ i \in \lbrace 1, \ldots, N \rbrace $ ‡ partir des observations $ \bz (t) $.

PlaÁons nous ‡ un instant fixe $ \tau $. Puis, pour chaque observation $ z_i $, on note $ \bw_{z_i, \tau} = \cW_ {z_i} (\bs, \tau) $ le vecteur ligne contenant les valeurs de la transformÈe en ondelettes ‡ l'instant $\tau$ pour un vecteur d'Èchelles $ \bs $ (dont la taille est notÈe $ M_s $). Ensuite, tous ces vecteurs sont rassemblÈs dans une  matrice de taille $ N \times M_s $ notÈe $ \bw_{\bz, \tau} $ telle que $ \bw_{\bz, \tau} = \left (\bw_ {z_1, \tau }^T \cdots \bw_{z_N, \tau}^T \right)^T $. La mÍme notation est utilisÈe pour la transformÈe en ondelettes des sources $ \bw_{\by, \tau} $. On suppose que la matrice $ \bA (t) $ varie lentement vis-‡-vis des oscillations des signaux. On peut alors montrer que la relation linÈaire~\eqref {eq:bss.model} devient (sous cette hypothËse) une relation entre les transformÈes en ondelettes de $ \by $ et $ \bz $ de la forme
\begin{equation}
\label{eq:approx.bss}
\bw_{\bz,\tau} \approx \bA(\tau)\bw_{\by,\tau}\ .
\end{equation}
Le thÈorËme suivant (o˘ $ \bM ^{\circ 2} $ dÈsigne le carrÈ ÈlÈment par ÈlÈment d'une matrice $ \bM $) donne une borne quantitative sur l'erreur d'approximation dans l'Èquation~\eqref{eq:approx.bss}.
\vspace{-1mm}
\begin{theorem}
Avec les notations ci-dessus, soit $ \bepsilon_{\tau} \in \CC^{N \times M_s} $ l'erreur d'approximation dans le domaine des ondelettes, dÈfinie comme suit:
\vspace{-1mm}
\begin{equation}
\bepsilon_{\tau} = \bw_{\bz,\tau} - \bA(\tau)\bw_{\by,\tau}\ . \\[-1mm]
\end{equation}
Supposons que les sources stationnaires sous-jacentes $ X_i $ ($ i = 1, \ldots, N $) soient des processus alÈatoires stationnaires au second ordre, de moyenne nulle, de puissance $ \sigma_X ^ 2 $. Alors l'erreur d'approximation $ \bepsilon _ {\tau} $ est une matrice alÈatoire complexe du second ordre, de moyenne nulle et circulaire. De plus, la variance des coefficients de la matrice d'erreur est bornÈe comme suit:
\begin{equation}
\EE\left\lbrace|\bepsilon_{\tau}|^{\circ 2}\right\rbrace \leq \sigma_X^2 k_\psi^2 \bA^{\prime \,\circ 2}_\infty\bgamma'_\infty (q^{3\bs})^T\ ,\\[-1mm]
\end{equation}
o˘
\vspace{-1mm}
\begin{align*}
k_\psi\in\RR_+:&& k_\psi &= \int_\RR |t\,\psi(t)| dt\ , \\
\bA'_\infty\in\RR^{N\times N}_+:&& (\bA'_\infty)_{ij} &= \sup_t |\bA_{ij}'(t)|\ ,\\
\bgamma'_\infty\in\RR^{N\times 1}_+: && (\bgamma'_\infty)_{i} &= \sup_t |\gamma_{i}'(t)|\ .
\end{align*}
\end{theorem}
\vspace{-1mm}
La dÈmonstration repose sur les mÍmes arguments que celle du thÈorËme~1 dans~\cite{Meynard18spectral}. Un dÈveloppement de Taylor de $ \bA $ en $ \tau $ permet la construction de la borne donnÈe ci-dessus.

Notons que, mis ‡ part les termes contrÙlant la borne sur l'erreur dans~\eqref{eq:approx.wavelet}, c'est-‡-dire la rÈsolution temporelle de l'ondelette et les variations des fonctions de dÈformation temporelle, la borne sur l'erreur dans~\eqref{eq:approx.bss} est Ègalement contrÙlÈe par les variations des coefficients de la matrice de mÈlange.

\vspace{-2mm}
\section{ProcÈdure d'estimation}
\vspace{-1mm}
La procÈdure d'estimation repose sur les relations approchÈes~\eqref {eq:approx.wavelet} et~\eqref{eq:approx.bss}, que nous supposons valides, c'est-‡-dire que $ \bA $ et $ \bgamma'$ varient suffisamment lentement. Cette hypothËse nous permet d'Ècrire une vraisemblance approchÈe sur les transformÈes en ondelettes des observations dans le cas gaussien. L'estimation s'appuie sur des transformÈes en ondelettes discrËtes, les paramËtres Èvoluant dans le temps sont donc estimÈs sur une grille de temps discrËte $ D $. Dans ce qui suit, nous dÈcrivons la procÈdure d'estimation pour un $ \tau \in D $ donnÈ. Par souci de simplicitÈ, nous introduisons les notations suivantes: $\bB_\tau = \bA(\tau)^{-1}$, $\theta_{i,\tau} = \log_q\left(\gamma'_i(\tau)\right)$ et $\btheta_\tau = (\theta_{1,\tau} \cdots \theta_{N,\tau})^T$.

\vspace{-2mm}
\subsection{Cadre probabiliste}
\vspace{-1mm}
DorÈnavant, les $ X_i $ sont supposÈs gaussiens. Il dÈcoule de cette hypothËse que la transformÈe en ondelettes de la $ i $\up{e} source $ Y_i $ est une matrice alÈatoire gaussienne, complexe circulaire (cf. la Proposition 3 dans~\cite{Omer17time}). Alors $\bw_{y_i,\tau}\sim \mathcal{N}(\mathbf{0},\bSigma_i(\theta_{i,\tau}))$, o˘
\[
\left[\bSigma_i(\theta_{i,\tau})\right]_{kk'}\!\! = q^{\frac{s_k+s_{k'}}{2}}\!\!\!\int_\RR\!\!\ccS_{X_i}(q^{-\theta_{i,\tau}}\xi)\overline{\hat\psi}\left(q^{s_k}\xi\right)\hat\psi\left(q^{s_{k'}}\xi\right) d\xi.
\]
Soit $ p_V $, de maniËre gÈnÈrique, la densitÈ de probabilitÈ d'une variable alÈatoire $ V $. Alors, l'hypothËse d'indÈpendance des sources conduit ‡ l'opposÈe de la log-vraisemblance suivante:
\begin{align*}
\ell_\tau(\bB_\tau,\btheta_\tau) \defeq& -\log(p_{\bw_{z,\tau}|(\bB_\tau,\btheta_\tau)}(\bw_{\bz,\tau};\bB_\tau,\btheta_\tau)) + c \\[-2mm]
 =& -\!M_s \log|\!\det(\bB_\tau)| \!+\!\dfrac1{2}\!\sum_{i=1}^N\log\left|\det\bSigma_i(\theta_{i,\tau}))\right| \\[-1mm]
&+\dfrac1{2}\sum_{i=1}^N[\bB_\tau\bw_{\bz,\tau}]_{i\cdot}\bSigma_i(\theta_{i,\tau})^{-1}[\bB_\tau\bw_{\bz,\tau}]_{i\cdot}^H\ ,
\end{align*}
o˘ $[\bM]_{i\cdot}$ indique la $ i $\up{e} ligne de la matrice $ \bM $, et $ \bM^H $ sa transposÈe conjuguÈe. Les estimations du maximum de vraisemblance (MV), c'est-‡-dire les minimiseurs de $ \ell_\tau (\bB_\tau, \btheta_\tau) $, peuvent Ítre ÈvaluÈs numÈriquement pour chaque $ \tau $, ils sont notÈs respectivement $ \tilde\bB_\tau$ et $ \tilde \btheta_\tau $.

\vspace{-2mm}
\subsection{Algorithme d'estimation}
\vspace{-2mm}
La stratÈgie d'estimation est d'alterner les estimations de $ \bB_\tau $, $ \btheta_ \tau $ et des spectres. L'algorithme~\ref{alg:estimation} (nommÈ JEFAS-BSS) synthÈtise toutes les Ètapes d'estimation. En ce qui concerne l'initialisation de la BSS, nous utilisons le rÈsultat de p-SOBI qui fournit une mÈthode de base donnant une matrice de mÈlange initiale variant dans le temps. L'algorithme JEFAS (qui est dÈtaillÈ dans~\cite {Meynard18spectral}) permet l'estimation des dÈformations temporelles et des spectres. Remarquons Ègalement que l'estimation de $\bB_\tau$ s'effectue avec un pas temporel de $\Delta_\tau$. La sÈparation s'effectue alors en considÈrant $\bB_\tau$ constante sur l'intervalle $ I_\tau = [\tau- \Delta_\tau / 2, \ \tau + \Delta_\tau / 2 [$.

\begin{algorithm}[t]
\floatname{algorithm}{Algorithme}
\legende{JEFAS-BSS}
\label{alg:estimation}
\begin{algorithmic}

\STATE {\bf Initialisation:} Evaluer $\tilde\bB^{(0)}_\tau$ au moyen de p-SOBI. Calculer les sources estimÈes $\tilde\by^{(0)}(\tau) = \tilde\bB^{(0)}_\tau\bz(\tau)$.

\STATE $\bullet$ $k \leftarrow 1$

\WHILE{le critËre d'arrÍt~\eqref{eq:stop.crit} est faux \AND $k\leq k_{max}$}

\STATE $\bullet$ Pour $i=1,\ldots,N$, estimer les paramËtres $\tilde\theta^{(k)}_{i,\tau},\ \forall\tau\in D$ et le spectre $\tilde\ccS_{X_i}^{(k)}$ en appliquant JEFAS ‡ $\tilde y_i^{(k-1)}$.

\FOR{$\tau=0,\Delta_\tau,\ldots,T$}

\STATE $\bullet$ Estimer $\tilde\bB^{(k)}_\tau$ par MV en remplaÁant $\btheta_\tau$ et $\ccS_X$ par leurs estimations actuelles $\tilde\btheta_\tau^{(k)}$ et $\left\lbrace\tilde\ccS_{X_i}^{(k)}\right\rbrace_{i=1,\ldots,N}$.

\ENDFOR

\STATE $\bullet$ Estimer les sources $\tilde\by^{(k)}(\tau) = \tilde\bB_\tau^{(k)}z(\tau)$.

\STATE $\bullet$ $k\leftarrow k+1$

\ENDWHILE
\end{algorithmic}
\end{algorithm}

Enfin, la convergence est contrÙlÈe en utilisant le rapport source ‡ interfÈrence (SIR) introduit dans~\cite{Vincent06performance}. Pour une source estimÈe donnÈe $ \tilde y_i $, le SIR, notÈ $ \text {SIR}(\tilde y_i, \by) $, quantifie la prÈsence d'interfÈrences dans $\tilde y_i$ provenant des autres sources $ y_j $, $ j \neq i $. Moins il y a d'interfÈrences, plus le SIR est grand. Comme nous n'avons pas accËs aux vÈritables sources, nous utilisons comme critËre d'arrÍt le SIR entre $ \tilde \by ^ {(k-1)} $ et $ \tilde \by^{(k)} $ (au lieu de $ \by $) qui donne une Èvaluation de la mise ‡ jour de la BSS et constitue donc une Èvaluation de convergence pertinente. Pratiquement, le critËre de convergence est testÈ sur le SIR moyennÈ entre les sources. Il est dÈfini comme suit:
\begin{equation}
\dfrac1{N}\sum_{i=1}^N\text{SIR}\left(\tilde y^{(k-1)}_i, \tilde \by^{(k)}\right) > \Lambda\ ,
\label{eq:stop.crit}
\end{equation}
o˘ $\Lambda$ est choisi suffisamment grand pour que l'amÈlioration de la qualitÈ de la BSS entre la $ (k-1) $\up{e} la $k$\up{e} itÈration de JEFAS-BSS ne soit pas significative.

\vspace{-1mm}
\section{RÈsultats}
\begin{figure}
\includegraphics[width=0.48\textwidth]{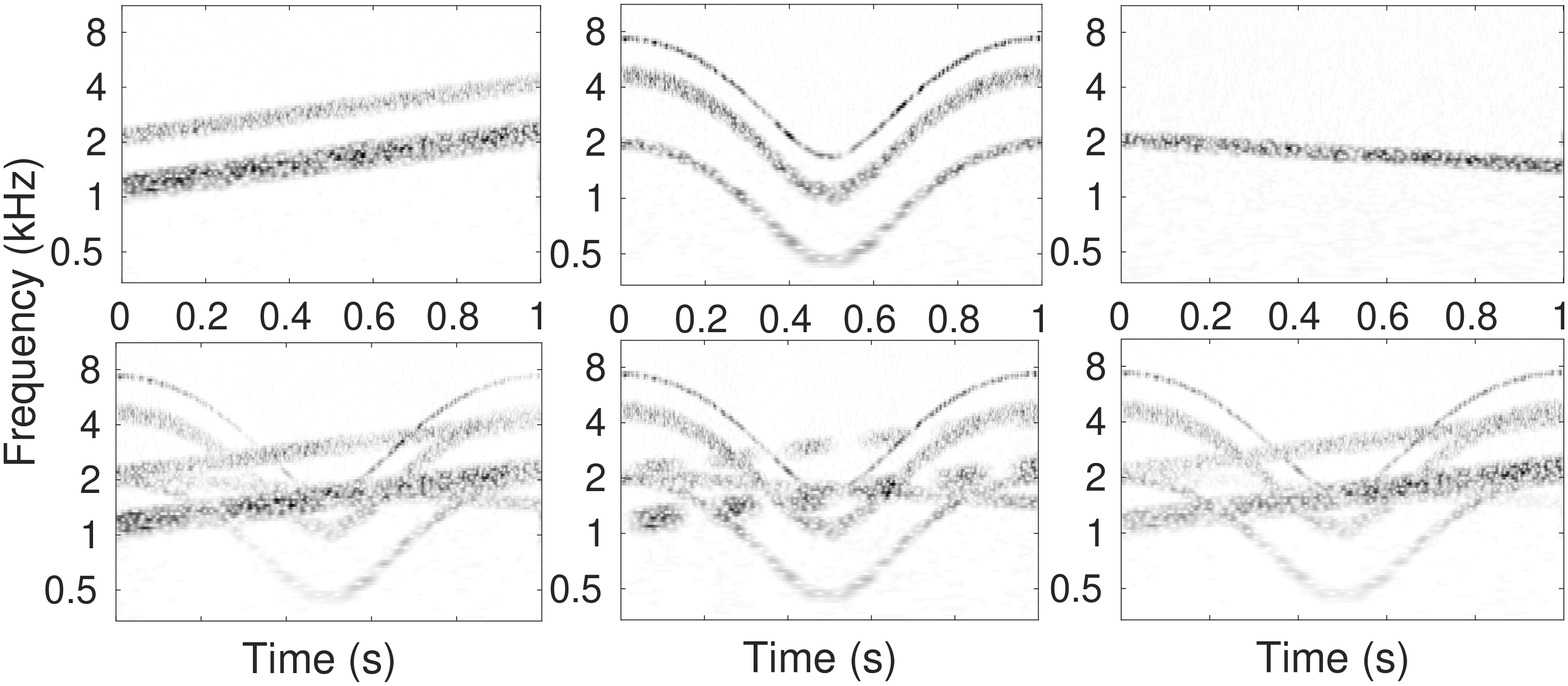}
\vspace{-8mm}
\legende{En haut: scalogrammes des trois sources non stationnaires. En bas: scalogrammes des trois observations.}
\label{fig:synthetic.mix}
\end{figure}

\vspace{-1mm}
Nous construisons un exemple synthÈtique avec $ N = 3 $ pour Èvaluer les performances de l'algorithme JEFAS-BSS. La durÈe des signaux est de 1 seconde et ils sont ÈchantillonnÈs ‡ $ F_s = 44,1 $~kHz. Les deux sources gaussiennes sont non stationnaires selon le modËle~\eqref{eq:nonstat.model}, leurs spectres de puissance sous-jacents $\ccS_{X_i}$ sont constituÈs de diffÈrentes fenÍtres de Hann ne se chevauchant pas. Les coefficients de la matrice de mÈlange Èvoluent sinusoÔdalement dans le temps (avec des frÈquences diffÈrentes). Les transformÈes en ondelettes des sources et des observations sont affichÈes sur la figure~\ref{fig:synthetic.mix}. Leurs supports Ètant superposÈs, un algorithme de BSS s'appuyant sur la parcimonie d'une reprÈsentation des signaux ne serait pas efficace ici.

\begin{figure}
\includegraphics[width=0.48\textwidth]{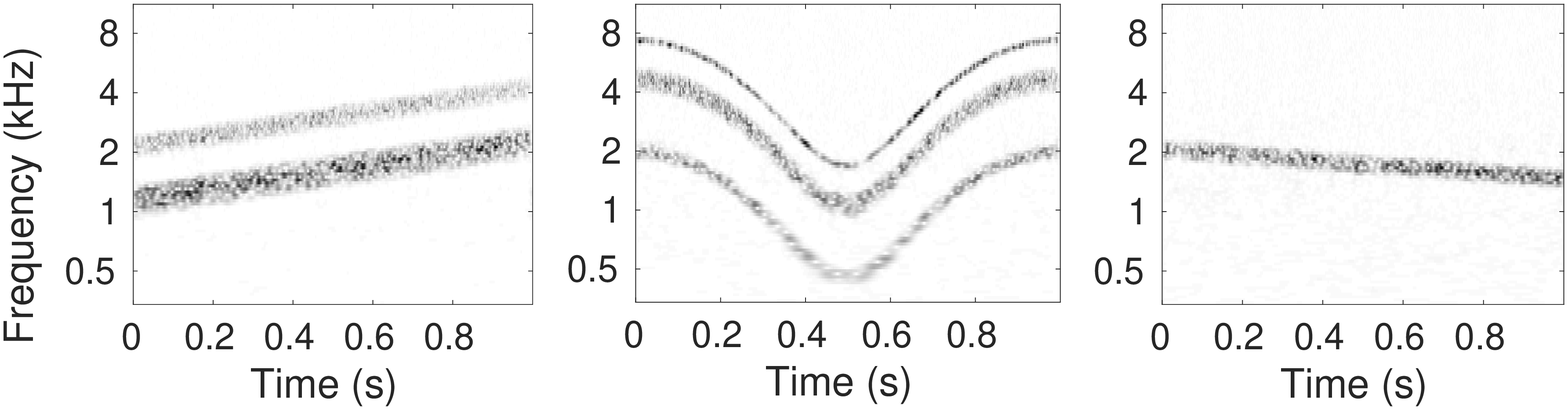}
\vspace{-8mm}
\legende{Scalogrammes des trois sources estimÈes.}
\label{fig:est.sources}
\end{figure}

\vspace{-3mm}
JEFAS-BSS est appliquÈ aux observations, il converge en 5~itÈrations. Les scalogrammes des sources estimÈes sont affichÈs sur la figure~\ref{fig:est.sources}. Par manque de place, les fonctions de dÈformation temporelle et les spectres estimÈs par JEFAS ne sont pas tracÈs ici. Pour une Ètude complËte des performances de JEFAS, nous renvoyons le lecteur ‡~\cite{Meynard18spectral}. Dans ce qui suit, nous nous concentrons sur les performances de la BSS.

Afin d'Èvaluer la BSS obtenue via JEFAS-BSS, nous la comparons ‡ d'autres algorithmes de BSS (SOBI, p-SOBI, QTF-BSS). Nous appliquons ces algorithmes ‡ 20 rÈalisations de l'exemple synthÈtique ci-dessus. Pour Èvaluer la qualitÈ des algorithmes BSS, nous calculons le SIR entre les vraies sources et leurs estimations en moyenne sur 20 simulations. Les rÈsultats sont donnÈs dans le tableau~\ref{tab:quality} ainsi que les Ècarts types (ET) correspondants. NÈanmoins, cette quantitÈ est un indicateur global de la qualitÈ de la BSS. Afin de suivre l'Èvolution temporelle de la qualitÈ de la BSS, nous introduisons l'indice d'interfÈrence inter-symboles normalisÈ $\rho$ (introduit dans~\cite{Moreau94one} pour l'indice non normalisÈ).
Cet indice est d'autant plus faible que la matrice $ \tilde \bB(t)\bA(t) $ est proche de l'identitÈ. De plus, on a toujours $\rho(t) \in [0,1] $. Par manque de place, l'Èvolution de l'indice d'Amari (en dÈcibels) pour les algorithmes BSS que nous Èvaluons n'est pas tracÈ. Cependant, les moyennes et Ècarts-types temporels de $\rho(t)$ sont mesurÈs. Les valeurs moyennes obtenues sur 20 simulations sont donnÈes dans le tableau~\ref{tab:quality}.
\begin{table}[t]
\legende{Comparaison du SIR et de l'indice d'Amari moyen pour quatre algorithmes de BSS.}
\centering
\begin{tabular}{|l||c|c||c|c|}
  \hline
   \multirow{2}{*}{Algorithme} & \multicolumn{2}{c||}{SIR (dB)} & \multicolumn{2}{c|}{$\rho$ moyen (dB)}\\
   \cline{2-5}
      & Moyenne & ET & Moyenne & ET\\
   \hhline{|=#=|=#=|=|}
   SOBI & $12,14$ & $3,59$ &$-6,62$ & $0,82$ \\
   \hline
   p-SOBI & $3,56$ & $1,75$ &$-9,03$ & $0,23$ \\
   \hline
   QTF-BSS & $0,69$ & $3,94$ & $-3,42$ & $0,42$ \\
   \hline
   JEFAS-BSS & $30,82$ & $0.63$ & $-15,47$ & $0,58$ \\
   \hline
\end{tabular}
\label{tab:quality}
\end{table}

Comme on pouvait s'y attendre, l'indice d'Amari et le SIR montrent clairement que JEFAS-BSS offre une meilleure BSS que les autres algorithmes. p-SOBI permet d'amÈliorer l'indice d'Amari par rapport ‡ SOBI mais le SIR est altÈrÈ en raison des effets de bords dus ‡ la segmentation de la BSS. Plus surprenant, les performances QTF-BSS sont infÈrieures ‡ celles de SOBI. Cela peut Ítre d˚ au fait que les estimations QTF-BSS et SOBI reposent sur un modËle matriciel de mÈlange constant bien que QTF-BSS soit adaptÈ ‡ la BSS de signaux non stationnaires, contrairement ‡ SOBI.

\vspace{-1mm}
\section{Conclusion}
\vspace{-1mm}
Dans cet article, nous avons prÈsentÈ JEFAS-BSS, un algorithme pour la BSS de mÈlanges instantanÈs non stationnaires d'une classe de signaux non stationnaires. JEFAS-BSS a ÈtÈ ÈvaluÈ sur un exemple synthÈtique sur lequel il surpasse nettement les mÈthodes de BSS existantes. JEFAS-BSS a Ègalement appliquÈ ‡ un mÈlange de signaux audios rÈels. Les rÈsultats numÈriques de cet exemple sont disponibles en ligne\footnotemark[1].

\bibliographystyle{ieeetr-fr}
\bibliography{Sampta17}

\end{document}